\documentstyle[12pt,epsf]{article}
\catcode`\@=11 \@addtoreset{equation}{section}  
\renewcommand{\theequation}                     
         {\arabic{section}.\arabic{equation}}   
\title{The quantum state-dependent gauge fields of Jacobi}
\author{P. Leifer}
\date{Crimea Federal University by the name of V.I. Vernadskii, \\
4 Vernadskii Av., 295007 Simferopol, Crimea, Russia}
\begin{document}
\maketitle
\begin{abstract}
It is commonly understood that the Yang-Mills non-Abelian gauge fields is the natural generalization of the well known Abelian gauge group symmetry $U(1)$ in the electrodynamics. Taking into account that the problems of the localization and divergences in QFT are not solved in the framework of the Standard Model (SM), I proposed a different approach to the quantum theory of the single self-interacting electron. In connection with this theory, I would like attract the attention to the state-dependent gauge transformations $U(1) \times U(N-1)$ associated with the Jacobi vector fields of the geodesic variations in the complex projective Hilbert space $CP(N-1)$ of the unlocated quantum states (UQS's).
\end{abstract}

PASC: 03.65 Pm, 03.65 Ca
\section{Introduction}
The extension of $U(1)$ gauge symmetry up to $U(1) \times SU(2)$ in the SM was very successful.  Nevertheless, even in the framework of the SM this symmetry leads to the artificial technical problems like anomalies \cite{Weinberg}. Furthermore, there is a requirement of the mathematically correct proof of the existence of such gauge fields in 4D with a mass gap \cite{YM}. Roughly speaking, the formulation of the Millennium problem in Yang-Mills gauge fields with a mass gap reflects the hopes of the mathematician and physicist majority. These are based on the standard picture of the Universe:  pseudo-rimannianan 4D spacetime in vicinity of stars with the pseudo-Euclidian limits from the cosmic to the sub-nuclear distances. To my mind, this program is intrinsically contradictable since such a model of the spacetime cannot serve as the fundamental element of the quantum field dynamics. New primordial elements are needed even in the theory of the single self-interacting quantum electron \cite{Le11,Le13,Le15}.
I think that we should follow the ideas of Einstein and Schr\"odinger in quantum physics as well as ideas of von Staudt in the projective geometry; namely: deleting the accent on the observation in quantum physics as well as deleting the metric relations in the projective geometry. Some aspects of such approach will be discussed in the present work.

The theory of a single self-interacting quantum object should solve problems left outside of the standard QFT and SM. This theory, being established, should solve the old problem of the rest mass and the correct form-factor of the surrounding electromagnetic-like ``field-shell". Therefore, free Dirac's electron and the free EM field may be treated as some ``border configurations" for some unified ``intermediate" self-interacting field. Geometrically, the non-normalizable plane wave solutions of the Maxwell and Dirac equations take the place of the indefinitely remote points of the Hilbert state space. Formally, these solution are ordinary points in the projective Hilbert state space $CP(\infty)$. There is some argument helping reduce the infinite dimension to small finite number.

Analysis of the foundations of quantum theory and relativity shows that it is impossible to use macroscopic primordial elements like material points (particles), solid scales and isochronises clocks trying to build consistent quantum theory. Even spacetime cannot save its independent and a priori structure \cite{Le13}. Therefore the unification of relativity and quantum principles may be formalized only in terms of new primordial elements. I proposed to use unlocated quantum states (UQS's), local dynamical variables (LDV's) corresponding pure quantum degrees of freedom as such primordial elements
and the geometric classification of their motions \cite{Le98,Le04,Le07,Le11,Le13,Le15}. So, the rays of the UQS's will be used instead of material points (particles) and the complex projective Hilbert state space $CP(N-1)$ where these states are moving under the action of the unitary group $SU(N)$ will be used instead of the spacetime motion. Thereby, the  \emph{distance between bodies will be replaced by the distance between unlocated quantum states} \cite{Le13,Le15}. The parameter $\tau$ which I called the ``quantum proper time" is the measure (in seconds) of the distance between unlocated quantum states concern omnipresent spin, charge, hypercharge, etc.

The self-interacting field configuration of quantum electron will be originated in $CP(3)$ \cite{Le11,Le13}. This field was taken in the form of the classical electromagnetic interaction term giving the rate of the unlocated quantum state (UQS) variation
$\frac{d \pi^i}{d \tau}=\frac{c}{\hbar}P^{\mu}(x)\Phi_{\mu}^i(\pi)$. It is assumed that one does not have a priori given spacetime environment but that the vector field of the lump-like potential of energy-momentum $P^{\mu}(x)$ taking the role of the placeholder in the dynamical spacetime (DST) \cite{Le11,Le13,Le15}. The coefficient functions $\Phi_{\mu}^i(\pi)$ of the $SU(4)$ generators corresponding to $\gamma$-matrices of Dirac \cite{Le97,Le98,Le13} take the place of non-Abelian spin/charge fields.

The decisive step from the quantum description in existing spacetime to the dynamics of UQS's in $CP(N-1)$ and the future lift in the frame fibre bundle where local DST arises in the section, leads to new symmetry and new conservation law. Namely, affine parallel transport of the ``proper" momentum of UQS $p^i=\frac{d \pi^i}{d \tau}$ is treated now as quantum inertia law \cite{Le13,Le15}. Therefore, the motion of UQS along a geodesic in $CP(N-1)$ is the simplest and fundamental motion to be represented by the non-linear field dynamics in DST realizing the ``inverse representation". The system of four quasi-linear PDE's describing $P^{\mu}(x)$ distribution in the state-dependent Lorentz frame is too complicated and its general solution is unknown but some particular solution has been found \cite{Le15}. In attempt to find the evident form of $P^{\mu}(x)$ I made some special assumption about the equation for arbitrary scalar function of the interval $f(s^2)$. This function leads to the scalar potential $P^0(x)$ oscillating between Coulomb and quantum potentials
\begin{eqnarray}\label{}
P^0=mc^2\frac{1+J_1(\frac{\alpha c t}{\sqrt 2} \sqrt{1-\beta^2})}{e \sqrt{1-\beta^2}},
\end{eqnarray}
which may be rewritten for the electron energy as follows:
\begin{eqnarray}\label{}
E=eP^0=\frac{mc^2}{\sqrt{1-\beta^2}} + \frac{e^2}{r_e} \frac{J_1(\frac{\alpha c t}{\sqrt 2} \sqrt{1-\beta^2})}{ \sqrt{1-\beta^2}},
\end{eqnarray}
where $\alpha=\frac{e^2}{\hbar c}$, $r_e=\frac{e^2}{mc^2}$, $\beta=\frac{r}{c t}$.
If one compares this solution with the total radial force
\begin{eqnarray}\label{}
\bar{f}^r_Q(\rho)+\bar{f}^r_V(\rho)=\frac{2}{\rho^3}-\frac{2}{\rho^2}
\end{eqnarray}
(10.3.7) of Yang \cite{Yang} one sees that two envelops of the maxima and the minima of this oscillating solution are similar to the quantum and Coulomb forces. This means that the stationary repulsive Coulomb potential and the attractive quantum force (playing the role of the ``Poincar\'e strength") may be treated as the stationary approximations to the oscillating self-potential generated by the motion of the spin/charge degrees of freedom. However, the arbitrary assumption that the scalar field $f(s^2)$ obeys
the  ``Lorentz-radial" form of the Klein-Gordon equation, insists us to find more firm foundations for studying the energy-momentum distribution.

The variation of the functional vector field $p^i$ gives some equation for unified self-interacting field. The independent element of the variation is a whole extremal (the geodesic in $CP(3)$ currying the unified field configuration), neither spacetime coordinates of the bare point-wise electron nor free EM potentials since these objects simply do not exist in this problem.

The most simple variations of the geodesic motions of the spin/charge QDF's is the geodesic variations. These variations have been described by the Jacobi equations which may have the Hamiltonian form for the so-called ``secondary Hamiltonian" and ``secondary extremals" in terms of Young \cite{Young}. Solutions of the Jacobi equations in $CP(N-1)$ are well known \cite{Besse} and may be associated with the isotropy group $H=U(1) \times U(N-1)$ taking the place of the gauge group in the new quantum field dynamics (QFD).

It worse while to recall that such mathematical objects as ``Jacobi magnetic field", ``Sasakian magnetic field" and ``K\"ahler magnetic field" are widely investigated by T. Adachi and coauthors \cite{Adachi1,Adachi2}. These objects were introduced as a generalization of the 2-form associated with the classical magnetic field. The quantum dynamics of UQS's in the $CP(N-1)$ being mapped into the frame fiber bundle by the structure group $H=U(1) \times U(N-1)$ leads to new field quasi-linear PDE's. Some their properties  will be represented in this work.

\section{Intrinsic unification of relativity and quantum principles. New primordial elements. The quantum principle of inertia}
I will discuss initially the attempt of the unification of relativity and quantum principles
undertaken by Yang in the framework of so-called Complex Mechanics (CM) \cite{Yang}.
There is some inconsistency in the relativistic CM proposed by Yang. His attempt to establish the peaceful coexistence between relativity and quantum physics can be accepted with serious reservations. Indeed, let me return to the derivation of the Klein-Gordon equation (12.2.15) under the assumption (12.2.12) (I will use the (xx.yy.zz) form of numeration of formulas according to \cite{Yang}).
Yang uses coordinates $q^{\alpha}(\tau)$  as coordinates of movable material points obey Hamilton equations controlling by an external potential and the IQP. Hence, he treats the Hamilton equations (12.2.16) and (12.2.20a), (12.2.20b) as equations of characteristics for the Klein-Gordon equation (12.2.15). But only straight lines (12.2.20a) serve as characteristics of the ordinary linear Klein-Gordon equation.

Strictly speaking neither Klien-Gordon equation nor modified non-linear Klien-Gordon equation discussed below are not the Hamilton-Jacobi equation since the last one should be of the first order in partial derivatives \cite{Courant,Arnold} and the Hamiltonian should be state-independent. In the case of state-dependent Hamiltonian (12.2.18) that literally reads as follows
\begin{eqnarray}
H(p(\tau),\Psi(q(\tau))) &=& \frac{\hbar^2}{2 m_0 \Psi}\{ [-(\frac{\partial \Psi}{c \partial t })^2+ (\frac{\partial \Psi}{ \partial x })^2+(\frac{\partial \Psi}{ \partial y })^2 +(\frac{\partial \Psi}{ \partial z })^2](1+\frac{1}{\Psi}) \cr &+& \frac{1}{\Psi} \Box \Psi\}
\end{eqnarray}\label{}
one has more complicated picture. Namely, the formal derivation of the Hamilton-Jacobi equation reads as a non-linear Klien-Gordon and the Hamilton equations cannot be treated as the characteristics equations for 1-jets system. In order to see it one should rewrite the action (12.2.12)  in the form
\begin{eqnarray}
S(\tau,\Psi(q^{\alpha}(\tau)))=W(\Psi(q^{\alpha}(\tau)))-H_0 \tau
\end{eqnarray}\label{}
if one takes into account the substitution (12.2.14)
\begin{eqnarray}
W(\Psi(q^{\alpha}(\tau)))=-i \hbar \ln \Psi(q^{\alpha}(\tau)).
\end{eqnarray}\label{}
This means that the Hamiltonian is state-dependent and the speed of the action variation (energy) is dynamical (not stationary). Then, using the chain rule one has
\begin{eqnarray}
\frac{\partial S}{\partial \tau} = \frac{\partial W}{\partial \Psi}
\frac{\partial \Psi}{\partial q^{\alpha}}
\frac{d q^{\alpha}}{d \tau}-H_0=
\frac{\partial W}{\partial \Psi}
\frac{\partial \Psi}{\partial q^{\alpha}}
\frac{d q^{\alpha}}{d \tau}-\frac{m_0c^2}{2}
\end{eqnarray}\label{}
and, therefore, we have following nonlinear equation
\begin{eqnarray}
\Box \Psi +2i\hbar m_0\frac{\nabla_{\mu}\Psi}{\Psi}\dot{x^{\mu}}+m^2_0c^2\Psi = 0
\end{eqnarray}\label{}
instead of the ordinary linear Klein-Gordon equation (12.2.15) since state-dependent Hamiltonian definitely could not lead to a linear wave equation.
In this case impossible to find separable solution of this modified Klein-Gordon equation in Yang's manner $\Psi(t,x,y,z)=T(t)X(x)Y(y)Z(z)$. Therefore, all successive procedure is formal for ad hoc choice of the entanglement with different wights of eigen-states of the ordinary Klein-Gordon equation.

Generally, state-dependent Hamiltonian of Yang $H(p(\tau),\Psi(q(\tau)))$ through the intrinsic quantum potential (IQP) leads to more general kind of the field equations than quasi-linear first order PDE's \cite{Courant,Arnold}.

One of the most controversial result is the ``corrected" formula (12.5.4) of Einstein for the kinetic energy of the relativistic particle
\begin{eqnarray}
E=\pm \frac{mc^2}{\sqrt{1-\frac{v^2}{c^2}}} \sqrt{1-\frac{2Q}{mc^2}}
\end{eqnarray}
This formula has been obtained from the expression (12.5.3)
\begin{eqnarray}
\frac{v^2}{c^2}= \frac{\dot{x}^2+\dot{y}^2+\dot{z}^2}{c^2}=\frac{\hbar^2}{p_0^2} (k_1^2+k_2^2+k_3^2)=\frac{k_0^2-m_0^2c^4}{E^2}.
\end{eqnarray}
where the last equality is incorrect.
It can be seen since $E=cp_0$ and, hence,
\begin{eqnarray}
\frac{\hbar^2}{p_0^2} (k_1^2+k_2^2+k_3^2)=\frac{\hbar^2c^2}{E^2} (k_1^2+k_2^2+k_3^2) \cr =\frac{c^2}{E^2} (p_1^2+p_2^2+p_3^2).
\end{eqnarray}
The last term was equalized by Yang as follows
\begin{eqnarray}
\frac{c^2}{E^2} (p_1^2+p_2^2+p_3^2)=\frac{k_0^2-m_0^2c^4}{E^2}.
\end{eqnarray}
However, the both equality $c^2(p_1^2+p_2^2+p_3^2)=k_0^2-m_0^2c^4=const$ and the ``new conservation law" $E^2+2m_0c^2 Q =k_0^2 =const$ (12.4.20), contradict to special relativity whereas experiments with pions, at least in the scattering regime, strongly support the Einstein's dispersion law.
All written above means that the superposition (entangled) state (12.4.1) cannot serve as a model of the real quantum particle. But one may note that even ordinary Klein-Gordon equation faces with serious difficulties of the physical interpretation \cite{Comay}.

Complicated motion of the model ``free" quantum particle under the action of the quantum potential in the CM \cite{Yang} is of course not a new phenomenon in physics. The most known is so-called Zitterbewegung found by Schr\"odinger during investigation of the Dirac equation for a free electron \cite{Schr}. These effects have a different nature but both of them pose the problem of the \emph{quantum inertial motion}. Is such motion does exist or not? If yes, then the geometric counterpart for quantum inertial motion should be changed. Indeed, in Newton/ Einstein mechanics inertial motion of the material point geometrically expressed by the straight line in the Euclidian space/ Minkowski spacetime. In general relativity indefinitely small mass moves along geodesic line in the Riemannian spacetime. Obviously, it is not of our case. Say, in the case of free Dirac's electron momentum is a constant since its operator commutes with the Hamiltonian but a velocity is an oscillating value. We cannot, therefore, appeal to straight line in spacetime as a geometric counterpart of the free electron motion. The situation in the CM even more complicated since momentum of the free point-wise quantum particle generally is not a constant.
Therefore, one should find some different geometric i.e. invariant counterpart for inertial quantum motion. Such invariants do exist but they are not connected directly to spacetime \cite{Le07,Le11}.

It is worse while to recall that Einstein sharply posed the question about the physical status of the spacetime \cite{Einstein_17}. The answer on this deep question is not clear up to now. The genesis of such concept is however well known. The class of the inertial systems contains (by a convention) the one unique inertial system - the system of distant stars. Then, on the abstract mathematical level arose a ``physical space" - the linear space with appropriate vector operations on forces, momenta, velocities, etc. General relativity and new astronomical observations concerning accelerated expansion of Universe show that all these constructions are only a good approximation, at best.

I seriously use the one of the main idea of Einstein about Gauss reference frame modeled by a deformable ``mollusc" \cite{Einstein_17}. But instead of the classical deformable body I use deformable quantum state since it has a clear geometric and group mechanism of the conservation and breakdown of the unitary symmetry.

The Galileo-Newton inertia principle tacitly assumes that there is some absolute global reference frame associated with the system of distant stars. This point of view looks as absolutely necessary for the classical formulation of the inertial principle itself. This fundamental principle has been formulated, say, ``externally", i.e. as if one looks on some massive body perfectly isolated from the Universe. In such approach only ``mechanical" state of relative motion of the body has been taken into account.
Einstein, however, clearly understood the logical inconsistence of the classical formulation of the inertia principle:
``\textbf{The weakness of the principle of inertia lies in this, that it involves an argument in a circle: a mass moves without acceleration if it is sufficiently far from other bodies; we know that it is sufficiently far from other bodies only by the fact that it moves without acceleration}" \cite{Einstein_21}. This argument may be repeated with striking force being applied to such non-localizable objects as plane waves of free particles since for such objects the ``sufficiently far" distance does not defined.
The sharp contradiction with classical formulation of the inertia principle gives QCD with the phenomenon of the asymptotic freedom of quarks. These massive objects directly break our classical understanding of inertia principle due to a new reality of the strong interaction if of course these objects are something more realistic then mathematical concepts of the theoretical scheme. This means that inertial motion is relative in a good approximation for ideally isolated bodies. But already the Newton's example of rotating bucket with water shows that there is an absolute motion since the water has a concave shape in any reference frame. Here we are very close to different - ``internal" formulation of the inertia principle and, probably, to understanding of the quantum nature of inertial mass. Namely, the ``absolute motion" should be turned  towards not outward, to distant stars, but inward -- to the state of deformable body.

In fact one should take into account that a force not only change the inertial character of its motion:  motion with the constant velocity transforms to accelerated motion; moreover -- the body deforms. Two aspects of the force action - acceleration relative inertial reference frame and deformation of the body are very important already on the classical level as it has been shown by Newton's bucket rotation.
In quantum physics internal state and its deformation esquires an imperative character:
since the second aspect of the force action -- a body deformation leads to the change of the  body state. In fact it is already a \emph{different body}, with different temperature, etc., \cite{Le11}. According to physical intuition one has in the case of the inertial motion an opposite situation -- the internal state of the body does not change, i.e. body is self-identical during inertial motion. It is tacitly assumed that all classical objects (frequently represented by material points) are self-identical and they can not disappear because of the energy-momentum
conservation law. The inertia law of Galileo-Newton ascertains this self-conservation ``externally". This means that objectively
\emph{physical state of body (temporary in somewhat indefinite sense) does
not depend on the choice of the inertial reference frame}. One may accept this statement as an ``internal" formulation of the inertial law for the quantum state that should be of course formulated mathematically. We put here some plausible reasonings leading to such a formulation.

Interaction generally leads to the absolute change (deformation) of the quantum
state \cite{Le97} since the quantum state is the state of motion.
Thereby, \emph{deformable unlocated quantum state} replaces the \emph{deformable body} of the classical physics and the group classification of its motion gives the natural geometric counterpart for quantum interaction or self-interaction.
Such approach leads to essentially new epistemology of the quantum theory: \emph{existence instead of observation is the main element of the new quantum paradigm.}

We will distinguish the ``total quantum state" $|\Psi(x,P,\pi)>$ in fiber bundle over $CP(N-1)$ and the internal unlocated quantum state $(\pi^1,...,\pi^{N-1})$ of pure QDF in the base manifold $CP(N-1)$. This construction is similar to very interesting approach of Marlow \cite{Marlow} with obvious reservations in physical interpretation of the mathematical formalism. An existence of a quantum object means some stability of so-called ``total quantum state". The complex analysis developed by Yang gives the consistent tools for the stability expression. One may note, that stabilization of the ``total quantum state" may be achieved due to the internal dynamics of the QDF. Hence, coordinates $(\pi^1,...,\pi^{N-1})$ of the flexible internal quantum state play the role of the ``shape" coordinates like it is in the flexible n-body gauge theory \cite{Littlejohn}.

Pure unlocated quantum states represented by the rays (points of complex projective Hilbert space $CP(N-1)$) will be used thereafter as fundamental physical concept instead of ``material point". It is assumed that quantum lumps (resembling Einstein's mollusc  \cite{Einstein_17}) obtained as solutions of some fundamental quantum non-linear relativistic wave equations will describe real relativistic quantum particles. Two simple observations may serve as the basis of the intrinsic unification of relativity and quantum principles.

The first observation concerns interference of quantum amplitudes (``total quantum states" $|\Psi(x,P,\pi)>$) in a fixed quantum setup.

A. The linear interference of quantum amplitudes shows the symmetries
relative space-time transformations of whole setup. This interference has
been studied in ``standard" quantum theory. Such symmetries reflects, say,
the \emph{first order of relativity}: the physics is same if any
\emph{complete setup} subject (kinematical,
not dynamical!) shifts, rotations, boosts as whole in single Minkowski
spacetime. According to our notes, to the transformations given above, one should add a freely falling quantum setup in gravitation field of a star.

The second observation concerns a dynamical ``deformation" of some quantum setup.

B. If one dynamically changes the setup configuration or its ``environment", then the amplitude of an event will be generally changed \cite{Berry1}. Nevertheless there is a different type of tacitly assumed symmetry that may be formulated on the intuitive level as the invariance of physical properties of ``quantum particles",
i.e. the invariance of their quantum numbers like mass, spin, charge,
etc., relative variation of quantum amplitudes. Such symmetry requires more general kind of state-dependent gauge transformations than typically used in gauge
theories. Physically it means that properties of, say, electrons are the same in two different setups $S_1$ and $S_2$.

\emph{One may postulate that the invariant content of this physical properties may be kept if one makes the infinitesimal variation of some
``flexible quantum setup"  reached by small variation of some
fields by adjustment of tuning devices.} This principle being applied to quantum system in pure quantum state was called the ``super-relativity principle".

There is an essential technical approach capable to connect super-relativity and the quantum inertia law. Namely, a new concept of the local dynamical variable (LDV) \cite{Le04} should be introduced for the realization of infinitesimal variation of a ``flexible quantum setup". Our construction is naturally connected with the  geometric phases \cite{Berry1} but we seek the local conservation laws for LDV's in the state space $CP(N-1)$. Following note is very important. In quantum physics the practical necessity requires so-called ``the second particle" \cite{AA88} playing the role of the external reference frame in the spacetime.
Despite of the statement of authors, such environment is not a ``quantum mechanical" but the classical too. Nevertheless, if one treats the concept of unlocated quantum state as an objective property and follows its continuous unitary evolution then it is possible to refer the ``present" state to the infinitesimally close ``previous" state in the Cartan spirit. Thereby, the concept of the deformable quantum state, LDV's and flexible reference frame rid us from the necessity in such external reference frame as the ``the second particle" \cite{Le13}.

\section{Functional Prespacetime and Dynamical spacetime}
Does electron use CGS in the calculations of the interaction energy?
How one may build spacetime geometry with help only quantum tools? What kind of the physical placeholder should be used? In classical physics one may use idealized material points, solid scale and a clock, but in quantum physics these are not appropriate tools.
If one has only single stable quantum particle like electron, how one should build some local spacetime geometry at least in its vicinity?

In the absence of the classical solid scales, four components of the vector field of the proper energy-momentum $P^{\mu}=(\frac{\hbar \omega}{c} ,\hbar \vec{k})$ will be used. This means that the period
$T$ and the wave length $\lambda$ of the oscillations associating with an electron's
field are identified with flexible (state-dependent) scales in the DST. Notice,
one may connect with any quantum particle sever natural state-dependent scales of the spacetime distance. If a mass $M$ is treated as an ersatz of the internal quantum motions
then in the ultra-relativistic limit of the de Broglie wavelength $L_{dB}=\frac{\hbar}{Mv}$ is the Compton wavelength $L_C=\frac{\hbar}{Mc}$. On the other hand, a self-field hypothesis of a charged particle leads to the so-called ``classical radius" $r_0=\frac{e^2}{Mc^2}$. Their ratio gives the fine structure constant $\alpha=\frac{r_0}{L_C}=\frac{e^2}{\hbar c}$ which is universal in the sense of independence on the lepton's mass. Dynamic behavior of the fundamental leptons is however quite different: electron is stable but muon and tauon are not. There is some deep reason for such difference. It looks like stable and unstable modes of some fundamental field dynamics.

The quantum field theory of a single self-interacting particle requires a new construction of the state space instead of the Fock space. I will introduce two new construction: functional prespacetime (FPST) and dynamical spacetime (DST). The representation of the QDF's dynamics along geodesic in $CP(N-1)$ should be realized in FPST by the field dynamics of the energy-momentum $P^{\mu}(x)$ in the DST.

Since we do not have the ``second particle" \cite{AA88} as an external reference frame, one may use only infinitesimally close ``previous" unlocated quantum state in the sense of Cartan.
Thereby, it is natural to take the energy-momentum field $P^{\mu}(x)$ as a lump-like placeholder in dynamical spacetime (DST). The UQS deviation is simply the distance in the Fubini-Study metric in $CP(N-1)$ (see below) but the calculation of the energy-momentum field $P^{\mu}(x)$ variation is complicated. This variation has a ``mechanical" part and electromagnetic-like potential generated by the motion of the spin/charge QDF.

The distance between two UQS's in $CP(N-1)$ may be measured by the interval in the Fubibi-Study metric $dS^2_{F.-S.}=G_{ik*}d\pi^i d\pi^{k*}$. Then the speed of the interval variation is given by the equation
\begin{eqnarray}
(\frac{dS_{F.-S.}}{d\tau})^2=G_{ik*}\frac{d\pi^i}{d\tau}\frac{d\pi^{k*}}{d\tau}
=\frac{c^2}{\hbar^2}G_{ik*}(\Phi^i_{\mu}P^{\mu})(\Phi^{k*}_{\nu}P^{\nu*})
\end{eqnarray}\label{115}
relative ``quantum proper time" $\tau$ where energy-momentum vector field $P^{\mu}(x)$ obeys field equations that will be derived later.
If one takes the equation for geodesic traversing with the constant speed $g=\sqrt{|f^1|^2+|f^2|^2+|f^3|^2}$
\begin{eqnarray}
\pi^1=\frac{f^1}{g}\tan g\tau,
\pi^2=\frac{f^2}{g}\tan g\tau,
\pi^3=\frac{f^3}{g}\tan g\tau
\end{eqnarray}\label{}
then the velocity has the components
\begin{eqnarray}
\frac{d\pi^1}{d\tau}=v^1=f^1(1+\tan^2 g\tau),
\frac{d\pi^i}{d\tau}=v^2=f^2(1+\tan^2 g\tau),
\frac{d\pi^i}{d\tau}=v^3=f^3(1+\tan^2 g\tau),
\end{eqnarray}\label{}
such as
\begin{eqnarray}
(\frac{dS_{F.-S.}}{d\tau})^2=G_{ik*}\frac{d\pi^i}{d\tau}\frac{d\pi^{k*}}{d\tau}
=g^2.
\end{eqnarray}\label{115}
The motion along the geodesic in $CP(N-1)$
with a variable speed (frequency) requires the variable energy-momentum field in order to support such a ``stationary" \emph{inertial} regime.

The geodesic flow in the projective Hilbert space will be generated by the Lagrangian ${\mathcal{L}}(x,P,\pi)=G_{ik*}\frac{d\pi^1}{d \tau} \frac{d\pi^{k*}}{d \tau}$. This is the base of the  dynamical picture of the self-interacting electron, then the canonical momentum is as follows
\begin{eqnarray}
p_j = \frac{\partial \mathcal{L}}{\partial (\frac{d\pi^j}{d\tau})}=G_{ik*}
\delta^i_j \frac{d\pi^{k*}}{d\tau} = G_{jk*}\frac{d\pi^{k*}}{d\tau} \cr
p_{j*} = \frac{\partial \mathcal{L}}{\partial (\frac{d\pi^{j*}}{d\tau})}=G_{ik*}
\delta^{k*}_{j*} \frac{d\pi^{i}}{d\tau} = G_{ij*}\frac{d\pi^{i}}{d\tau}.
\end{eqnarray}\label{33}
Therefore
\begin{eqnarray}
p^{j*} = G^{j*s}G_{sk*}
\frac{d\pi^{k*}}{d\tau} = \delta^{j*}_{k*}\frac{d\pi^{k*}}{d\tau} = \frac{d\pi^{j*}}{d\tau},
\end{eqnarray}\label{34}
and $p^i = \frac{d\pi^i}{d\tau}$.

The canonical self-energy momentum $p^i$ of the relativistic electron will be expressed by the contraction of unknown proper 4-potential $P^{\mu}=p^{\mu} - \frac{e}{c}A^{\mu}$ and the coefficient functions $\Phi^i_{\mu}$ of the $SU(4)$ generators
\begin{eqnarray}
\vec{\gamma_{\mu}}=\Phi ^i_{\mu}\frac{\partial }{\partial \pi^i} + c.c.
\end{eqnarray}
forming the state-dependent tetrad in the \emph{functional prespacetime} (FPST) that connected with spin/charge dynamics in $CP(3)$
\begin{eqnarray}
p^i = \frac{d \pi^i}{d \tau}=\frac{c}{\hbar}P^{\mu}(x)\Phi^i_{\mu}(\pi).
\end{eqnarray}
The connection of this FPST manifold with the dark matter concept is the problem of the future investigations.
Such internal quantum spin/charge dynamics cannot be directly connected with the spacetime dynamics of electron since the first one assumes dynamics relative quantum proper time $\tau$ that measures of the $CP(3)$ distance between unlocated (in spacetime) quantum states $\pi^{i'} =\pi^i+ d\pi^i$ and $\pi^i$ along geodesic so that
\begin{eqnarray}
\frac{d \pi^i}{d \tau}
&=&c(\frac{p^{\mu}}{\hbar}-\frac{e A^{\mu}}{\hbar c})\Phi^i_{\mu}\cr
&=&c(\frac{p^{\mu}}{\hbar}-\frac{e^2 \kappa^{\mu}}{\hbar c})\Phi^i_{\mu}=c(\kappa^{\mu}_{(0)}-\alpha  \kappa^{\mu}_{(1)})\Phi^i_{\mu}\cr
&=&(\Omega^{\mu}_{(0)} - \alpha \Omega^{\mu}_{(1)})\Phi^i_{\mu},
\end{eqnarray}
whereas the spacetime dynamics of electron connected with motion of the ``field shell" wrapping spin/charge degrees of freedom in local dynamical spacetime (DST). Here $\kappa^{\mu}$ denotes the 4-field of the ``wave vector" with physical dimension $m^{-1}$ decomposed in the series in the fine structure constant $\alpha=\frac{e^2}{\hbar c }$.

The variation of the energy-momentum of the self-field revealed due to the separation of the ``Lorentz reference frame motion" from the internal motion of QDF gives the propagation of the quantum particle. The measure of ``propagation" will be defined by the distribution of the proper energy-momentum of the self-field ${\bf P}=P^{\mu}{\bf e}_{\mu}=(\frac{\hbar \omega}{c}{\bf e}_{0} + \hbar \vec{k}{\bf e}_{\vec{x}})$ where local frame in DST ${\bf e}_{\mu}=\frac{\partial }{\partial x^{\mu}}$ will be treated as an accelerated ``observer". Therefore, the covariant derivative
\begin{eqnarray}
P^{\nu}_{;\mu} = \frac{\partial
P^{\nu}}{\partial x^{\mu} } + \Gamma^{\nu}_{\mu \lambda}P^{\lambda}
\end{eqnarray}
arises during the infinitesimal Lorentz frame transformations where it is assumed that
the state-dependent metric tensor $g_{\mu \nu}$ generated by the internal motions of the spin/charge QDF may be represented up to the terms of the $O(|x^{\alpha}|^2)$ as follows:
\begin{eqnarray}
g_{0 0} &=& 1+2a_{\alpha}x^{\alpha},\cr
g_{0 \gamma} &=& 2\epsilon_{\gamma \alpha \beta}x^{\alpha} \omega^{\beta},\cr
g_{\alpha \beta} &=& -\delta_{\alpha \beta}, \quad (\alpha, \beta, \gamma = 1,2,3).
\end{eqnarray}\label{115}
The general coordinate invariance do not considered
in this paper since the spacetime curvature is the second order effect \cite{G}
in comparison with Coriolis contribution to the pseudo-metric
\cite{Littlejohn} in boosted and rotated state dependent ``Lorentz" reference frame.
The affine connection in this ``quantum Lorentz reference frame" is as follows:
\begin{eqnarray}\label{43}
\Gamma_{\nu 0}^{\mu}=-\Omega_{\nu}^{\mu}=-\left( \begin {array}{cccc}
0&a_1&a_2&a_3 \cr
a_1&0&-\omega_3&\omega_2 \cr
a_2&\omega_3&0&-\omega_1 \cr
a_3&-\omega_2&\omega_1&0
\end {array} \right),
\end{eqnarray}
or in the components
\begin{eqnarray}
\Gamma_{00}^0=\Gamma_{000}=0,\cr
\Gamma_{j0}^0=-\Gamma_{0j0}=+\Gamma_{j00}=+\Gamma_{00}^j=a^j,\cr
\Gamma_{k0}^j=\Gamma_{jl0}=-\omega^i \epsilon_{0ijk}
\end{eqnarray}
\cite{G}.

There is the natural geometric approach concerning conditions of the invariance and deformation of UQS's in terms of the $CP(N-1)$ dynamics \cite{Le97,Le13,Le15}. But this dynamics requires the ``inverse representation" of $CP(3)$ motions by the field of energy-momentum propagation in DST due to the separation
of the ``quantum Lorentz transformations" of the energy-momentum, i.e. the introduction of the horizontal and vertical components of the kinetic energy.
The vertical component arises from the ``Lorentz" transformations of the coordinates $x^{\mu}$ and the momentum $P^{\mu}$, whereas the horizontal component arises due to the motion of the spin/charge degrees of freedom.

The symbolic picture of the FPST and DST in the sections of the frame fiber bundle is depicted in Fig.1.

\vskip .1cm
\begin{figure}[ht]
\centerline{\epsfxsize=\hsize \epsffile{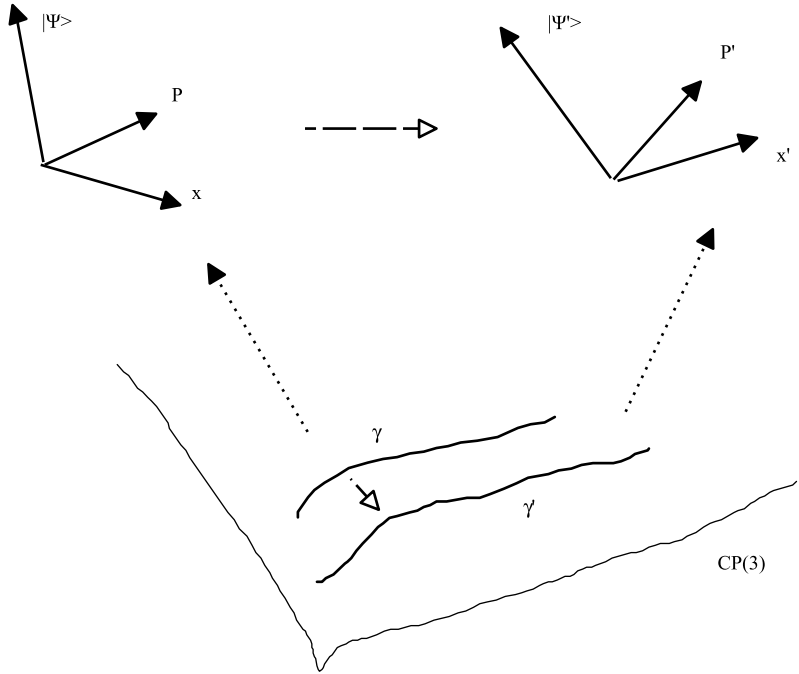}}
\caption{The symbolic picture of the frame fiber bundle over $CP(3)$.  The transformation of the geodesic to geodesic is lifted in this fiber bundle. The sections are given by the choice of the gauge that defines the local ``orientation" of the dynamical spacetime relative the functional prespacetime. Two field configurations in one of the section are shown.}
\label{fig.1}
\end{figure}
\vskip .1cm

Generally, the process of a measurement in physics has to be associated with the comparison process of some dynamical variable with corresponding scale. Newtonian physics established invariant character of a measurement in the sense of its independence on the choice of inertial reference frame (IRF). This assumption was formulated as \emph{the inertia principle of Galileo-Newton}.

Einstein, taking into account the finite speed of light and its invariance relative choice of the IRF found that result of the distant measurement depend on the choice of the IRF, i.e. so-called relativistic kinematics and dynamics replaced absolute character of a measurement of Newtonian mechanics. Meanwhile, general relativity already renders global spacetime coordinates in fact in physically senseless values.

Quantum theory brings a new kind of ``relativity" -- dependence of the result of a measurement on the apparatus used for the measurement.
This kind of relativity is so radical that the indeterministic paradigm and even agnostic philosophy takes over the habitant deterministic character of predictions in exact sciences. Deep difficulties in the ``standard" QFT almost insist to try to find new invariants, in fact a new quantum geometry based on the intrinsic properties of quantum states of the elementary particles \cite{Le97}. The most principle difference between the ``standard" QFT and the real situation in quantum physics is as follows.

Main experimental method of the investigation in high energy physics is scattering accelerated quantum particles. However all attempts to rich deep zone of the quantum particle with help of the outer ``probe" particles leads to creation of new particles as a consequence of new quantum degrees of freedom de-freezing.
But stable quantum particles and even unstable particles demonstrate temporal classical behavior in relatively weak external fields. Therefore their own field structure ``sweeps" all zone occupied by the quantum particle without dramatic multiple particle creations. Therefore, it is reasonable to assume that (internal) quantum degrees of freedom of quantum particle forming unlocated quantum state may be ``wrapped" in the ``field shell" that distributed in DST.
The general idea is that the local DST arises in a cross-section of the frame fiber bundle due to the state-dependent Lorentz frame embedding.
Thereby, we would like to save ordinary spacetime four dimension and local Lorenzian character of the metric.

The new principle of ``super-relativity", i.e. relativity to the choice of any conceivable quantum setup seems to be correct, since pure quantum degrees of freedom like charge, spin, hypercharge, etc. are anywhere the same \cite{Le13}. This principle makes accent on the objective invariant nature of the unlocated (omnipresent) quantum vacuum states of motion and operates with the notion of ``existence" instead of ``observation".
Namely, existence associated with the stability of self-interaction dynamics expressed in the terms of quantum state itself without any reference to external apparatus. Geometry of the projective Hilbert space takes the place of the fundamental geometry where gauge dynamics of the quantum degrees of freedom defines the basic properties of the ``elementary" quantum particles. The habitant global omnipresent spacetime should be replaced by the specific section of the fiber bundle over $CP(N-1)$. In the framework of the affine gauge dynamics the term ``existence" esquires the exact mathematical sense in relation to the stability of quasi-linear PDE's solutions.

The separation of the spacetime coordinates from pure quantum degrees of freedom may be associated with state-dependent Lorentz transformations of the proper energy-momentum. This fact leads to the new construction of the dynamical spacetime.
Namely, the spacetime distance occurs as a result of unholonomy of the state space of the unlocated quantum degrees of freedom. In this picture the global spacetime coordinates have no physical meaning at all (at least in the framework of the single particle problem). Only local coordinates relative state-dependent Lorentz reference frames may be defined due to more or less complicated operational procedure. This procedure is based on the introduction of the local state-dependent Lorentz reference frame ${\bf e}_{\mu}=\frac{\partial }{\partial x^{\mu}}$ of ``observer".

The dynamical spacetime (DST) is pure local construction built for description of energy-momentum distribution which I called the ``field-shell" of the quantum particles.
This spacetime is non-distinguishable from state-dependent Lorentz frame and moves together with it in a cross-section of the principle fiber bundle over $CP(N-1)$. There are two times in this theory. One time is the ordinary Einstein's time in the DST. The second one is the ``quantum proper time" that serves as a measure of the distance between two quantum states in the base state manifold $CP(N-1)$ expressed in seconds. I will use anywhere symbol $``t"$ for the Einstein time and the symbol $``\tau"$ for the quantum proper time.

\section{Hamiltonian vector field for the self-interacting electron}
The standard approach of the quantum theory is to take some classical Hamilton function and to ``quantize" it. If one starts, however, from the internal QDF's one does not have the classical analogy. Dirac already undertook some attempt to derive the relativistic equation for a fermion with two additional internal degrees of freedom \cite{Dirac_1971} besides the spin and the electric charge. These two degrees of freedom where modeled by two harmonic oscillators. It turns out that such a model cannot describe some physical particles since their self-potential is trivial.

The relativistic electron of Dirac has spin/charge degrees of freedom decoded in $\gamma$-matrices. I proposed more ``soft" model of the non-local electron where $\gamma$-matrices were replaced by the continuous complex vector fields on the $CP(3)$ of the $SU(4)$ generators corresponding to $\gamma$-matrices \cite{Le11}. Thereby, the spin/charge QDF's were ``dissolved" into three complex projective coordinates  $(\pi^1,\pi^2,\pi^3)$.
The operator of the energy-momentum in $CP(3)$
\begin{equation}\label{}
\vec{P}=i P^{\mu}(x) \Phi^i_{\mu}(\pi)\frac{\partial}{\partial \pi^i}+c.c.=i P^{\mu}(x)
\vec{\gamma}_{\mu}(\pi) + c.c.
\end{equation}
was used instead of the Dirac's operator of the energy-momentum
\begin{equation}\label{7}
\hat{p}= \hat{\gamma}^{\mu}p_{\mu} =i\hbar \hat{\gamma}^{\mu}\frac{\partial}{\partial x^{\mu}}
\end{equation}
that acts in the direct product $S=C^4 \times H_D$. Here $H_D$ means the Hilbert space where defined four differentiable functions $\psi^m$. Such direct product seems to be too ``stiff" and we try to find more flexible construction of the energy-momentum operator. Namely, more reasonable to work in the fibre bundle and only in a some section it is possible to have local splitting into ``horizontal=internal" and ``vertical=external" contribution of the energy. In such a case the mass is not a free phenomenological parameter but a state-dependent functional of the state. Then the energy of the total quantum state $|\Psi(x,P,\pi)>$ of the single quantum electron is the  sum of \emph{velocities of the action (phase) variations} from one UQS to another under the self-conservation of the electron.

Quantum inertia law in the case of inertial motion of the non-local
quantum electron may be formulated as the conservation of its internal \emph{momentum of unlocated quantum state} $p^i=\frac{d \pi^i}{d \tau}$ \cite{Le11,Le13,Le15}.
Thereby, the conservation law of the momentum vector field in $CP(3)$ during inertial evolution will be expressed by the affine parallel transport
in both spaces: the momentum $p^i$ in $CP(3)$ with the affine connection (...) and the energy-momentum $P^{\mu}(x)$ in the local DST, where coefficients of the affine connection in the accelerated ``quantum Lorentz reference frame" where given above (3.13).

The problem of the construction of the \emph{Hamiltonian vector field} for relativistic self-interacting lump-like electron may be treated as an opposite to the Aharonov-Anandan projection problem \cite{AA90} of the Hilbert space trajectories onto the close curve in the $CP(N-1)$. First of all, however, one needs to clarify the Aharonov-Anandan statement about the ``Hamiltonian independence" of the distance in $CP(N-1)$. This ``proof" is incorrect since the Hamiltonian from the ``infinite number of Hamiltonians which would evolve the state along a given curve $C$ in $\mathcal{P}$" should be path-dependent. The short phrase of authors means a lot of things:

1). There is some path $C$ in $CP(N-1)$ as a function of the parameter $\tau$ such that
$C: [0,\pi] \to (\pi^1(\tau),...,\pi^{N-1}(\tau))$.

2). In some variable basis $\{|0(t)>,|1(t)>,...,|a(t)>,..., |(N-1)(t)> \}$ in the Hilbert space $\mathcal{H}$ one has for some curve
\begin{eqnarray}
|\psi(t, \tau)>= \sum_0^{N-1} \psi^a(\tau)|a(t)> \cr = \psi^0|0(t)> + \pi^1(\tau)\psi^0|1(t)>+,...,+ \pi^{N-1}(\tau) \psi^0 |(N-1)(t)>.
\end{eqnarray}

3). The Hamiltonian which should be found is the path-dependent functional operator (functional vector field)$H=H(t,C)=H(t,\pi^1(\tau),...,\pi^{N-1}(\tau))$.

4). Projection of Aharonov-Anandan means that for a different Hamiltonian $H_1=H(t,C)_1=H_1(t,\pi^1(\tau),...,\pi^{N-1}(\tau))$ one has the different solution $|\phi(t)>$ in the Hilbert space $\mathcal{H}$, but the components of these solutions obey following equalities
$\pi^1(\tau)=\frac{\psi^1(\tau)}{\psi^0} =\frac{\phi^1(\tau)}{\phi^0} ,...,\pi^{N-1}=\frac{\psi^{N-1}(\tau)}{\psi^0} =\frac{\phi^{N-1}(\tau)}{\phi^0}$.

Hence, there is only the trivial difference between two solutions:
$|\phi>=(\phi^0 \frac{\psi^{0}}{\psi^0}, \phi^0 \frac{\psi^1(\tau)}{\psi^0},...,\phi^0\frac{\psi^{N-1}(\tau)}{\psi^0})=
\frac{\phi^0 } {\psi^0}|\psi>$
that is possible only if two Hamiltonians are proportional.

The class of the proportional Hamiltonians that provide the ``transport of the state along a given curve $C$ in $\mathcal{P}$" in such a formulation of the problem is really infinite but the statement about the independence of the measure $s$ on such a Hamiltonian is incorrect. One may see it, say, from the formula (15) in \cite{AA90} of the ``quantum-mechanical clock" motion. The formula
$s=\frac{2\eta}{\sqrt{3}\hbar}\sqrt{J^2+J}t$
clearly shows the dependence of $s$ on the energy $\eta$ which may variate from Hamiltonian to Hamiltonian. However, the distance in ${\mathcal{P}} = CP(N-1)$ where  \emph{a given curve} lies, is definitely independent on the choice of any Hamiltonian!
Furthermore, in this example it is unclear what is the projection line in $\mathcal{P}$. The statement of the authors is merely: ``the clock is not moving along a geodesic".
For us it is interesting the opposite problem: to find at least one Hamiltonian from this indefinite class for a given curve in $\mathcal{P}$, i.e. one needs
to lift dynamics along closed geodesic in ${\mathcal{P}} = CP(N-1)$ into the frame fiber bundle. Namely, one needs to find the Hamiltonian providing such kind of the spacetime field dynamics that the Hilbert space trajectory (the solution of the ``Schr\"odinger equation", see below) will be projected onto a closed geodesic of $CP(N-1)$.

The velocity of the total $|\Psi(x^{\mu},P^{\mu},\pi^i)>$ variation is given by the linear ``super-Dirac" equation
\begin{eqnarray}\label{35}
i \{\hbar V^{\mu}_Q \frac{\partial |\Psi>}{\partial x^{\mu}}  + \hbar [-V^{\mu}_Q  \Gamma^{\nu}_{\mu \lambda}P^{\lambda}-\frac{c}{\hbar}(\frac{\partial
\Phi_{\mu}^n (\pi)}{\partial \pi^n} \cr + \Gamma^m_{mn} \Phi_{\mu}^n(\pi) ) P^{\nu}P^{\mu}]\frac{\partial |\Psi>}{\partial P^{\nu}} \cr + c P^{\mu}\Phi_{\mu}^i(\pi)\frac{\partial |\Psi>}{\partial \pi^i} + c.c.\} = m(x,P,\pi) c^2 |\Psi>
\end{eqnarray}
that equivalent to the ``Schr\"odinger equation"
\begin{eqnarray}\label{36}
i \hbar \frac{d |\Psi >}{d \tau} = \vec{H}(x,P,\pi ) |\Psi > \cr
=\vec{U}|\Psi > + \vec{T}|\Psi >
= m(x,P,\pi) c^2 |\Psi >
\end{eqnarray}
for the single total state function $|\Psi(x^{\mu},P^{\mu},\pi^i)>$ of self-interacting quantum electron ``cum location" moving in DST like a free material point with the rest dynamical mass $m(x,P,\pi)$ and continuous spin/charge variable $(\pi^1,\pi^2,\pi^3)$.

One may treat the functional Hamiltonian vector field $\vec{H}(x,P,\pi)$ as an inertia operator
whose action may be gauge invariantly divided on the horizontal and the vertical components. The vertical component generated by the ``gradients" of the total wave function $|\Psi >$ in $x^{\mu}$ and $P^{\mu}$ due to quantum Lorentz transformations.
\begin{eqnarray}
|V_v> = \vec{U}|\Psi > =  \hbar (\frac{d x^{\mu}}{d \tau}\frac{\partial |\Psi >}{\partial x^{\mu}} + \frac{d P^{\mu}}{d \tau}\frac{\partial |\Psi >}{\partial P^{\mu}}).
\end{eqnarray}
The linear operator of the kinetic energy
\begin{eqnarray}
\vec{T} = \hbar p^i \frac{\partial}{\partial \pi^i} + c.c. = \hbar \frac{d \pi^i}{d \tau}\frac{\partial}{\partial \pi^i}+ c.c. \cr =cP^{\mu}(x) \Phi_{\mu}^i(\pi)\frac{\partial}{\partial \pi^i} + c.c. = cP^{\mu}(x) \vec{\gamma}_{\mu}
\end{eqnarray}
is similar to the Dirac operator
$\hat{T} =c \hat{\gamma}_{\mu}p^{\mu}=c\hat{\gamma}_{\mu}\frac{\partial }{\partial x^{\mu}}$ that generates the horizontal part of the kinetic energy
\begin{eqnarray}
|V_h> = \vec{T}|\Psi > = \hbar \frac{d \pi^i}{d \tau}\frac{\partial |\Psi >}{\partial \pi^i}+ c.c. \cr = cP^{\mu}(x) \Phi_{\mu}^i(\pi)\frac{\partial |\Psi >}{\partial \pi^i} + c.c.
\end{eqnarray}

This operator is the \emph{tangent vector field} to $CP(3)$ arising from the self-interaction term. This has been used in order to determinate the inertial (dynamical) mass and an additional potential terms. Let me return in connection with this problem to the field equations in the short form
\begin{eqnarray}\label{}
\Omega^{\mu}_{\nu}x^{\nu}\frac{\partial P^{\sigma}}{\partial x^{\mu}} -\Omega^{\sigma}_{\nu}P^{\nu}=0
\end{eqnarray}
where $\Omega^{\mu}_{\nu}$ is the matrix (3.12) whose elements are the state-dependent boosts and rotations of the co-moving Lorentz reference frame
\begin{eqnarray}
a_{\alpha} &=&c L_{\alpha}\frac{\hbar - \sqrt{\hbar^2+4P^0\hbar(L_1x+L_2y+L_3z)}}{2\hbar(L_1x+L_2y+L_3z)} \cr
\omega_{\alpha} &=& \frac{c \epsilon^{\beta}_{\alpha \gamma}L_{\beta} P^{\gamma}}{\hbar(1+\frac{a_1x+a_2y+a_3z}{c})}.
\end{eqnarray}
\cite{Le15}.
The partial solution
\begin{eqnarray}\label{}
P^{\mu}(x)=C_1 x^{\mu} + C_2 f(x_{\mu}x^{\mu})x^{\mu}+C_3g(x_{\mu}x^{\mu})\Omega^{\mu}_{\nu}x^{\nu}
\end{eqnarray}
has been recently found if the state-dependent (through the spin/charge current $L_{\mu}(\pi)$) $\Omega^{\mu}_{\nu}$ and $g_{\mu \nu}$ obey following non-covariant conditions playing the role of the gauge restrictions on the coordinates
\begin{eqnarray}\label{}
\frac{\partial \Omega^{\mu}_{\nu}}{\partial x^{\lambda}}x^{\nu}=0,\cr
\frac{\partial g_{\mu \nu}}{\partial x^{\lambda}}=\Gamma_{\mu, \nu \lambda}+\Gamma_{\nu, \mu \lambda}=0.
\end{eqnarray}
The energy component $P^0(x)$ may be found separately from the spatial components $P^{\alpha}(x)$. Namely, from the first algebraic equation of (4.10) one has
\begin{eqnarray}\label{}
(P^0(x)-f(x) x^{0})^2 =
= g^2(x) \frac{c^2}{4}(1-\sqrt{1+\frac{4P^0(xL_1+yL_2+zL_3)}{\hbar}})^2,
\end{eqnarray}
where it was assumed that the constants $C_1,C_2,C_3$ were absorbed by the redefined functions $f(x)=C_1+C_2f(x_{\mu}x^{\mu}), g(x)=C_3g(x_{\mu}x^{\mu})$. One of the four solutions is as follows
\begin{eqnarray}\label{}
P^0(x) = \frac{cg(x)}{2}+x^0f(x)+\frac{c^2g^2(x)(xL_1+yL_2+zL_3)}{2\hbar}\cr
+\frac{\sqrt{(4f(x)x^0c^2g^2(x)+2\hbar c^3g^3(x))(xL_1+yL_2+zL_3)+...+\hbar^2c^2g^2(x)
}}{2\hbar},
\end{eqnarray}
where the intermediate terms were omitted.
In this expression one should put $g(x)=2m(x)c$ in order to have ordinary relativistic decomposition of $P^0(x)=mc^2+...$. Three other solutions are similar to the ordinary Dirac states as follows: $|+mc^2,\uparrow >$  and $|-mc^2,\uparrow>$, $|-mc^2,\downarrow >$. The spatial components of the energy-momentum $P^{\alpha}(x)$ may be found from the linear non-homogeneous system
\begin{eqnarray}\label{}
(1+A_1)P^1+0+B_1P^3=R_1,\cr
B_2P^1+(1+A_2)P^2+0=R_2,\cr
0+B_3P^2+(1+A_3)P^3=R_3,
\end{eqnarray}
where
\begin{eqnarray}\label{}
A_{\alpha}=\frac{g(x) \epsilon_{\alpha \beta \gamma}x^{\beta} L_{\gamma}}{\hbar(1+\frac{a_1x+a_2y+a_3z}{c})}, \cr
B_{\alpha}=- \frac{g(x) \epsilon_{\alpha \beta \gamma}x^{\gamma} L_{\beta}}{\hbar(1+\frac{a_1x+a_2y+a_3z}{c})}, \cr
R_{\alpha}=x^{\alpha}f(x).
\end{eqnarray}
Thereby, one has
\begin{eqnarray}\label{}
P^1(x)=\frac{R_1(1+A_2+A_3)+A_2A_3R_1+B_1B_3R_2-A_2B_1R_3-B_1R_3}
{1+A_1+A_2+A_3+A_1A_2+A_1A_3+A_2A_3+A_1A_2A_3+B_1B_2B_3},\cr
P^2(x)=\frac{R_2(1+A_1+A_3)+A_1A_3R_2+B_1B_2R_3-A_3B_2R_1-B_2R_1}
{1+A_1+A_2+A_3+A_1A_2+A_1A_3+A_2A_3+A_1A_2A_3+B_1B_2B_3}, \cr
P^3(x)=\frac{R_3(1+A_1+A_2)+A_1A_2R_3+B_2B_3R_1-A_1B_3R_3-B_3R_2}
{1+A_1+A_2+A_3+A_1A_2+A_1A_3+A_2A_3+A_1A_2A_3+B_1B_2B_3}.
\end{eqnarray}

The trivial solutions of (4.9) with constant $f(x)=c_1, g(x)=c_2$ are regular at the origin   and they are formally possible but these solutions are physically unacceptable because of the divergences on the infinity. The definition of cut-off free functions $f(x), g(x)$ due to the restrictions (4.12) will be discussed elsewhere.

We discussed up to now only single electron as a dynamical process along geodesic $\gamma_1$. Second electron may be represented similarly
as a dynamical process along geodesic $\gamma_2$.
Two tangent vector fields of the geodesic flows may be transformed one to another by the isotropy group $H=U(1) \times (U-1)$ of some UQS since it acts on the symmetric,
homogeneous, totally geodesic state space $CP(N-1)$ playing the role of the base manifold. One-parameter evolution of UQS's in the projective Hilbert space $CP(N-1)$ assumes the lift of this dynamics in the ``field frame" $P^{\mu}(x)$ fiber bundle under the action of the structure group $H=U(1) \times U(N-1)$.  The potential part of the $P^{\mu}(x)$ field may be associated with the gauge field connecting two electrons where the second electron playing the role of the ``second particle" in the DST. These fields naturally relate to the Jacobi fields.

\section{ The gauge fields of Jacobi}
As far as we know the gauge and matter fields have common quantum dynamical variables corresponding to the internal quantum degrees of freedom, e.g. spin, charge, color, etc. One may assume that just UQS's of motions in $CP(N-1)$ comprise ``UQS's vacuum" and its excitations are concentrated lump-like solutions of the field equations corresponding quantum self-interacting particles. From this point of view, equations of pure EM fields and the Dirac equations for free electrons serve merely as equations for 4D ``boundary" field configurations for unknown equations of the lump-like excitations of the UQS's vacuum. Since just in the vicinity of an electron the linear QED should be improved, I will use the ``intermediate" expression for the self-interaction energy  which is a convolution of the
energy-momentum $cP^{\mu}(x)$ to be found, and the dimensionless spin/charge non-Abelian current vector field $\Phi^i_{\mu}(\pi)$ which is known \cite{Le13}.
Then the variation of this functional vector field (i.e. operator, not functional!) should give some equation for unified self-interacting field.
The most simple variations of the geodesic motions of the spin/charge QDF's is the geodesic variations. These variation locally have been described by the Jacobi equations which may have the Hamiltonian form for the so-called ``secondary Hamiltonian" and ``secondary extremals" in terms of Young \cite{Young}.

These transformations rotate geodesics passing some fixed point in $CP(N-1)$ as whole.  The infinitesimal geodesic variations may be described by the Jacobi equation
\begin{eqnarray}\label{}
\frac{\delta^2 J^A}{\delta \sigma^2}+R^A_{BCD}\frac{dX^B}{d \sigma}\frac{dX^C}{d \sigma}J^D = 0,
\end{eqnarray}
where real coordinates $X^A$ with the indexes $A,B,C,D = 1,2,...,2(N-1)$ are such that
$\pi^j = X^{2j-1}+i X^{2j}$.

I will use without essential loss of generality the partial geodesic $\gamma_0: \vec{\pi}(\sigma) = (\tan \omega \tau,0,...,0), \omega= \frac{\arctan \lambda_0}{\sigma_0} $
that connects two points of $CP(N-1)$ $\vec{\pi}(0)=(0,...,0)$ and $\vec{\pi}(\sigma_0)=(\lambda_0,...,0)$.
The basis $\xi^i_{(k)}$ parallel to the given geodesic $\gamma_0$ may be chosen
as follows:

\begin{eqnarray}\label{}
\xi^1_{(1)} &=& \frac{c_0}{\cos^2 \omega \tau}, \xi^i_{(1)} = 0, (i=2,3,..., N-1), \cr
\xi^i_{(k)} &=& 0, (i \neq k), \xi^i_{(k)} = \frac{c_0}{\cos \omega \tau} (i=k).
\end{eqnarray}

In such a basis the general form of the Jacobi field  may be represented
\begin{eqnarray}\label{}
J^A= N_0 f^B \xi^A_{(B)} + (N_1\tau+N_2) v^B \xi^A_{(B)}
\end{eqnarray}
where the term $N_0 f^B \xi^A_{(B)} = N_0 \sin \omega \tau \xi^A_{(B)} $ is the normal Jacobi field.

The geodesic rotation generates the gauge field in the DST and in opposite, the influence of the external gauge field leads to the geodesic variation.
Two close velocities in $CP(3)$ may serve as a boundary conditions and they should be agreed in order to define so-called ``the field of the generator" (operator of the UQS momentum (3.8)) of the variational problem for the geodesic family in $CP(3)$.
I try to get the evident form of the energy-momentum $P^{\mu}(x)$ as the consequence of this agreement. Such agreement may be expressed by the simplest physical equation of motion
\begin{eqnarray}
\nabla_{\dot{\gamma}}\dot{\gamma} = J, or \cr
\frac{\delta p^i}{\delta \tau}= J^i.
\end{eqnarray}\label{}
This is the obvious generalization of the
\begin{equation}
\nabla_{\dot{\gamma}}\dot{\gamma} =0,
\end{equation}
describing the inertial motion discussed above.

I will analyze the new equations (5.4) in attempt to separate the schedule of the UQS evolution in $\tau$ along some $CP(3)$ geodesic from the geodesic line itself. This means that we should find some motion with \emph{variable frequency} i.e. a trajectory defining by some Hamiltonian vector field. The Jacobi field (the normal plus longitudinal parts) has following components
\begin{eqnarray}
J^1=N_0 \frac{\tan \omega \tau}{\cos \omega \tau} + (N_1 \tau + N_2)p^1; \cr
J^2=N_0 \tan \omega \tau + (N_1 \tau + N_2)p^2; \cr
J^3=N_0 \tan \omega \tau + (N_1 \tau + N_2)p^3.
\end{eqnarray}\label{}
Therefore one has three equations for the UQS's momentum $p^i=\frac{d \pi^i}{d \tau}$
\begin{eqnarray}
\frac{d p^1}{d \tau}-\sin(2\omega \tau)(p^1)^2 -N_0\frac{\tan \omega \tau}{\cos \omega \tau} -(N_1 \tau + N_2)p^1 =0; \cr
\frac{d p^2}{d \tau}-\frac{\sin(2\omega \tau)}{2}p^1p^2 - N_0 \tan \omega \tau - (N_1 \tau + N_2)p^2 =0; \cr
\frac{d p^3}{d \tau}-\frac{\sin(2\omega \tau)}{2}p^1p^3 -N_0 \tan \omega \tau - (N_1 \tau + N_2)p^3 =0.
\end{eqnarray}\label{}

Some fundamental analytical solutions of these equations is as follows.

1. ``Free" solutions for the non-deformed geodesic motion ($N_0=N_1=N_2=0$)
\begin{eqnarray}
p^1 &=& \frac{\omega }{\cos^2 \omega \tau+C_1\omega}; \cr
p^2 &=& \frac{C_2}{\sqrt{\cos 2 \omega \tau +1+2C_1 \omega}}; \cr
p^3 &=& \frac{C_3}{\sqrt{\cos 2 \omega \tau +1+2C_1 \omega}}.
\end{eqnarray}\label{}

2. ``Longitudinal mode I" solutions ($N_0=N_1=0, N_2 \neq 0$)
\begin{eqnarray}
p^1 &=& \frac{(N_2^2+4\omega^2)e^{N_2 \tau} }{2\omega \cos (2 \omega \tau) e^{N_2 \tau}-N_2 \sin (2 \omega \tau)e^{N_2 \tau}+C_1(N_2^2+4\omega^2)};
\end{eqnarray}\label{}
the analytic solutions for $p^2, p^3$ are not achievable.

3. ``Longitudinal mode II" ($N_0=0, N_1 < 0, N_2 \neq 0$)
\begin{eqnarray}
p^1 &=& \frac{4\sqrt{N_1}e^{-\frac{N_1\tau^2}{2}+N_2\tau}} {\sqrt{2\pi}i[e^{\frac{(N_2+2i\omega)}{2N_1}}
erf(\frac{N_1\tau-N_2-2i\omega}{\sqrt{2N_1}})-e^{\frac{(N_2-2i\omega)}{2N_1}}
erf(\frac{N_1\tau-N_2+2i\omega}{\sqrt{2N_1}})]+4C\sqrt{N_1}};
\end{eqnarray}\label{}
the analytic solutions for $p^2, p^3$ are not achievable.

4. ``Normal mode" ($N_0 \neq 0, N_1 = N_2 = 0$)
\begin{eqnarray}
p^1=\frac{N}{D}, \cr
N =-\sqrt{-\cos \omega \tau}[\omega^2 I(1,\sqrt{-\frac{8\cos \omega \tau}{\omega^2}}) + C K(1,\sqrt{-\frac{8\cos \omega \tau}{\omega^2}})];\cr
D = \cos \omega \tau(-C\omega \sqrt{-\cos \omega \tau}
K(1,\sqrt{-\frac{8\cos \omega \tau}{\omega^2}})
+ \sqrt{2} C \cos \omega \tau K(0,\sqrt{-\frac{8\cos \omega \tau}{\omega^2}})\cr
- \omega^3 \sqrt{-\cos \omega \tau} I(1,\sqrt{-\frac{8\cos \omega \tau}{\omega^2}}) -\omega^2 \cos \omega \tau \sqrt{2} I(0,\sqrt{-\frac{8\cos \omega \tau}{\omega^2}}));
\end{eqnarray}\label{}
the analytic solutions for $p^2, p^3$ are not achievable.
These functions put definite restrictions on $P^{\mu}(x)$. This will be discussed in a future article.

\section{Conclusion}
The attempt of the intrinsic unification of the quantum and the relativity principles has been discussed. The gauge potential $P^{\mu}(x)$ being expressed in terms of ``quantum Lorentz" transformations was generated by changes in UQS under condition of vanishing the covariant derivative of the proper momentum $p^i$ of the UQS, i.e. the affine parallel transport in $CP(3)$. Convention of the attaching the ``quantum Lorentz" frame to the deformable UQS is a gauge convention. Then a change of the convention we shall regard as a gauge transformation.

The non-Abelian character of the gauge field was originated by the geometry of the spin/charge motion of the Dirac electron along geodesics of $CP(3)$. The velocity of the simplest deformations of such a motion are properly defined by the Jacobi equations. The definition of the free functions $f(x), g(x)$ including in the form-factor of the quantum electron is postponed for a future work.

\end{document}